\documentclass[jgrga]{AGUTeX}
\usepackage{graphicx}
\usepackage{amsmath}
\usepackage{amssymb}
\setkeys{Gin}{draft=false}
\authorrunninghead{STEVENSON, J. E. H. and PARNELL, C. E.}
\titlerunninghead{I. Nature of the reconnection}
\authoraddr{Corresponding author: J. E. H. Stevenson, School of Mathematics and Statistics, Mathematical Institute, North Haugh, St Andrews, Fife, KY16 9SS, Scotland (jm686@st-andrews.ac.uk)}
\begin{document}
\title{Spontaneous reconnection at a separator current layer.\\ I. Nature of the reconnection}
\authors{J. E. H. Stevenson,\altaffilmark{1} and C. E. Parnell\altaffilmark{1}}

\altaffiltext{1}{School of Mathematics and Statistics, Mathematical Institute, North Haugh, St Andrews, Fife, KY16 9SS, Scotland}

\begin{abstract}
Magnetic separators, which lie on the boundary between four topologically-distinct flux domains, are prime locations in three-dimensional magnetic fields for reconnection, especially in the magnetosphere between the planetary and interplanetary magnetic field and also in the solar atmosphere.
Little is known about the details of separator reconnection and so the aim of this paper, which is the first of two, is to study the properties of magnetic reconnection at a single separator. Three-dimensional, resistive magnetohydrodynamic numerical experiments are run to study separator reconnection starting from a magnetohydrostatic equilibrium which contains a twisted current layer along a single separator linking a pair of opposite-polarity null points. The resulting reconnection occurs in two phases. The first is short involving rapid-reconnection in which the current at the separator is reduced by a factor of around 2.3. Most ($75\%$) of the magnetic energy is converted during this phase, via Ohmic dissipation, directly into internal energy, with just $0.1\%$ going into kinetic energy. During this phase the reconnection occurs along most of the separator away from its ends (the nulls), but in an asymmetric manner which changes both spatially and temporally over time. The second phase is much longer and involves slow impulsive-bursty reconnection. Again Ohmic heating dominates over viscous damping. Here, the reconnection occurs in small localised bursts at random anywhere along the separator.
\end{abstract}

\begin{article}

\section{Introduction}
Many highly energetic space physics processes involve magnetic reconnection, such as solar and stellar flares, coronal mass ejections, interactions between planetary and interplanetary magnetic fields and substorms in magnetospheres, etc. Two key papers that explained the basics of generalised three-dimensional (3D) reconnection \citep{Schindler88,Hesse88} provide the corner stones to all 3D reconnection studies. 
In particular, one of their main results reveals that, unlike 2D reconnection, 3D reconnection can either occur with or without null points.

In 3D, reconnection has been shown to occur at a multitude of sites, such as at topological features including 3D null points (locations where all three components of the magnetic field equal zero) \citep[e.g.,][]{Craig95,pontin04a,pontin05b,Pontin07,priest09a,Masson09,pontin11a} and magnetic separators (special field lines that link pairs of 3D null points) \citep[e.g.,][]{Priest96,Longcope96,Longcope01,Haynes07,Parnell10a,Parnell10b,Wilmot11} or at the geometrical features known as quasi-separatrix layers (regions of magnetic field, which at one end are closely anchored, but at the other end are anchored far apart) \citep[e.g.,][]{Demoulin95,Demoulin96,Demoulin97,aulanier05a,Aulanier06} and in twisted or braided flux tubes \citep[e.g.,][]{Galsgaard96,GN97a,GN97,demoortel06a,demoortel06b,wilmot07a,Browning08,Hood09,Bareford13}. 

The reconnection that we focus on here is that associated with magnetic separators. Such reconnection has been invoked as an explanation for a number of specific observed solar flares \citep[e.g.,][]{Longcope05,Barnes05}, for flux emergence events \citep[e.g.,][]{Parnell10b,MacTaggart14} and in the heating of the quiet-Sun \citep[e.g.,][]{Close04}. Furthermore, magnetic separators have been found to occur on a wide range of scales for they are found in global magnetic field extrapolations from photospheric synoptic magnetograms extending over several solar radii \citep[e.g.,][]{Platten14,Edwards15a} and also in local magnetic extrapolations of the quiet-Sun from high-resolution magnetograms \citep[e.g.,][]{Close05}.  
Additionally, magnetic reconnection plays an important role in planetary magnetospheres enabling the interaction of the interplanetary and planetary magnetic fields and powering flux transfer events and substorms. Indeed, separators have been identified in numerous models of the Earth's magnetosphere \citep[e.g.,][]{Hu04,Laitinen06,Laitinen07,Dorelli07,Dorelli08,Dorelli09,Hu09,Pulkkinen10,Ouellette10,Peng10,Cnossen12,Komar13}. Notably, \cite{Komar13} have shown that on the dayside magnetopause, regardless of the direction of the interplanetary magnetic field (IMF), magnetic reconnection is most likely to occur at separators.

There are only limited ``observations'' of magnetic separators in the Earth's magnetosphere. This is not surprising since, as pointed out by \citet{Parnell10a},  ``for 3D magnetic fields, global 3D topological structures and local 3D field structures do not [necessarily] coincide''. Since separators are global structures this means it is difficult to determine separators unless the magnetic field is known ``practically everywhere, which is obviously not the case from solar and magnetospheric observations''. Nonetheless published results claiming to have detected reconnection at separators using Cluster data include the work of \citet{Phan06} who reported an observed separator (called, in their paper, an X-line) on the Earth's dayside, whilst \citet{Xiao07} report having observed separator reconnection in the nightside magnetosphere. \citet{Deng09} and \citet{Guo13} also report examples of separator reconnection observed using Cluster data.

We note that in some of the works referenced here, the term `X-line' is used in place of `magnetic separator' since it is thought that the projected magnetic field in a plane perpendicular to a 3D magnetic separator is X-type and looks like a 2D null point. This is not true since in 3D the global topology and local magnetic field are not necessarily coincident as they are in 2D. So although the projection of the global topology in a plane perpendicular to a separator is X-type, the projection of the local magnetic field in such a plane may be X-type or O-type \citep{Parnell10a}. Further to this, the term `X-line' is often used to describe a line of nulls, which is an unstable feature \citep{Hesse88} and entirely different from a magnetic separator. A 3D magnetic separator may be regarded as a 2.5D X-line (a 2D null point with a guide field).

A separator is a special field line that connects a positive null to a negative null (separators may also connect two bald patches or a null and a bald patch \citep[see, for example,][]{titov93,haynesPhD07}, but we do not concern ourselves with such separators here). A positive/negative 3D null is associated with a pair of spine lines (field lines which are directed into/away from the null point) and a separatrix surface (a surface of field lines pointing away from/into the null point). A separator is formed when; (i) the separatrix surfaces of two oppositely signed nulls intersect, (ii) the spine of one null point intersects with the separatrix surface of another null point of the same sign and (iii) the spines of two oppositely signed nulls intersect. (i) is the only one which is both general and generic since the others are unstable to perturbations (i.e., small perturbations would cause them to disappear resulting in a change to the structure of the topology of the magnetic field) and, hence, is the only type considered in this paper. A separator lies on the boundary of four topologically distinct flux domains (i.e., movement between the domains would result in a discontinuous jump in field line mapping) and so field lines in the vicinity of separators are sensitive to flows across these boundaries. Hence, currents build easily along separators \citep{LauFinn,Haynes07,Parnell10a,Parnell10b,Stevenson15} and so they are prime locations where 3D reconnection occurs. 

Separator reconnection has been studied both analytically and numerically \citep{Sonnerup79,LauFinn,Longcope96,GN97,G00,Longcope01,PC06,Haynes07,Dorelli08,Parnell08,Parnell10a,Parnell10b,Komar13}. Reconnection at separators is known to be different from 3D null point reconnection. In particular, \cite{Parnell10a} studied the dynamic nature of separator reconnection in a model where two nulls, positioned on the base of the box, were moved together via boundary driving such that their separatrix surfaces intersected to form separators. They found that the parallel electric field along a separator varies spatially and temporally and may be multiply peaked, implying that there can be one or more local ``hot spots'' of reconnection along a separator. Additionally, these reconnection hot spots, which coincided with counter-rotating flows, occur away from the ends of the separator making separator reconnection distinct from 3D null point reconnection. They discussed that the projected magnetic field in planes perpendicular to a separator may either be X-type or O-type, with, in their model, the X-type projected field regions corresponding to O-type flow and weaker reconnection and the O-type projected field regions corresponding to X-type flow and stronger reconnection. This finding has been questioned by some who believed the twisted nature of the field was a result of the specific driving in the experiment. 

In this work, we detail the properties of 3D separator reconnection in a non-driven experiment, to avoid any such problems, and also where the null points and separator are far from the boundary. The model starts from a MHS equilibrium, containing a separator current layer and excess energy above that of a potential field, formed through the non-resistive relaxation of an initially non-potential, non-force-free field, discussed in detail in \citet{Stevenson15}. Reconnection is triggered at the separator current layer using an anomalous diffusivity to mimic the onset of micro-instabilities. In nature it is believed that the slow driving of complex magnetic fields leads to equilibria forming which have current layers located, for instance, where the field line mapping is discontinuous. Reconnection can occur at such current layers via micro-instabilities once the length scales are sufficiently short such that the magnetic Reynolds number, $R_m \le 1$. 

This approach is different to many works, which start from potential (minimum energy) fields that are driven on the boundaries (at slow or fast rates) to induce reconnection. In these situations, the resulting reconnection rate is typically found to depend on the rate of driving \citep[e.g.,][]{Galsgaard05}, and the nature of the initial magnetic configuration.

The approach we use has been used before to investigate reconnection at 2D magnetic null points, e.g., \citet{FFP12,FFP12b} where high and low plasma-beta reconnection regimes were studied, respectively. The high and low-beta MHS configurations, from which the reconnection experiments start, involve enhanced current not just at the null point, where it forms a null current layer, but also along the separatrices of the null. Everywhere else the current is very low. In contrast to \citet{Longcope07} and \citet{Longcope12} (zero-beta models), the numerical models of \citet{FFP12} (high-beta), and \citet{FFP12b} (low-beta) find that the reconnection process converted most of the magnetic energy (stored in the 2D null current layer configuration) directly into internal energy, via Ohmic dissipation, with only a little being converted initially into kinetic energy and then damped due to viscosity. Additionally, \citet{FFP12} found that the value of the magnetic diffusivity affects not only the reconnection rate, but also the amount of magnetic energy converted into kinetic and internal energy. 

All experiments studied here start with equilibria involving a separator current layer embedded in a high-beta plasma (detailed in \citet{Stevenson15}). So called cluster separators (separators that link nulls clustered within a single weak field region \citep{Parnell10b}) are likely to be in a high-beta plasma. These separators are typically short (approximately 1-2 Mm in length) \citep{Parnell10b}. The vast majority of null points in the solar atmosphere occur low down \citep[e.g.,][]{Regnier08,Longcope09,Edwards15b} and thus will reside in the high-beta chromospheric region of the Sun; intercluster separators (linking nulls from different null clusters) that lie within the chromosphere are likely to have lengths below 10 Mm (considering the typical size of small-scale photospheric magnetic features). The plasma beta in the magnetosphere lies between 1 and 10 \citep{Trenchi2008}, thus separator reconnection here is also likely to occur in a high-beta plasma. The lengths of these separators have been found to be greater than 200 Mm \citep{Komar13,KomarPhD}. As a result of this wide range of separator lengths, the results that we present here are non-dimensional. We explain the appropriate scalings that should be applied to produce dimensional results applicable to the many varied space physics situations where reconnection occurs. 

Here, we study the nature of non-driven 3D separator reconnection at a generic single-separator current layer. Specifically, we choose an idealistic setup (a straight single separator) to allow us (i) to relatively easily analyse the nature of the reconnection found at all points along the separator and (ii) the nature of the waves and flows \citep{Stevenson15_jgrb} resulting from the reconnection. In particular, we do not attempt to simulate one specific space physics event, but rather produce a model whose general results may be applicable in many events involving separators. We look to answer: what is the partitioning of the magnetic energy released by the reconnection?; how is the reconnection rate affected by the value of the diffusivity, the size of the diffusion region and the background viscosity? We compare our results to those of \citet{Parnell10a}, e.g., is the reconnecting magnetic field elliptic (helical) or hyperbolic?; to what extent do the null points play a role in separator reconnection?; does reconnection occur along the entire length of the separator?

In Sect.~\ref{sec:equb} we detail the properties of the MHS equilibrium and the current layer it contains. The numerical model used to carry out the experiments is discussed in Sect.~\ref{sec:nummodel}, followed by an analysis of the energetics of the main experiment (Sect.~\ref{sec:energetics}). We then examine the nature of the reconnection (Sect.~\ref{sec:phase1}) by analysing where the reconnection is strongest and the plasma properties there. Finally, we detail the effects of varying the diffusivity, the size of the diffusion region and the background viscosity in the system (Sect.~\ref{sec:etavisc}) before summarising our results (Sect.~\ref{sec:conclusions}).

\section{MHS Equilibrium Current Layer}\label{sec:equb}
The MHS equilibrium used as the initial condition for our numerical experiments was formed through non-resistive MHD relaxation \citep{Stevenson15}. The initial equilibrium is identical to that found by \citet{Stevenson15} with the exception of an increased domain length in the $z$-direction, but the grid resolution is the same. The dimensions of the numerical domain, in Cartesian coordinates, are $-1.0 \le x,y \le 1.0$ and $-1.75 \le z \le 2.75$ and in grid cells is $(512,512,768)$.

\begin{figure}[h!]
\centerline{
\includegraphics[width=0.4\textwidth,clip]{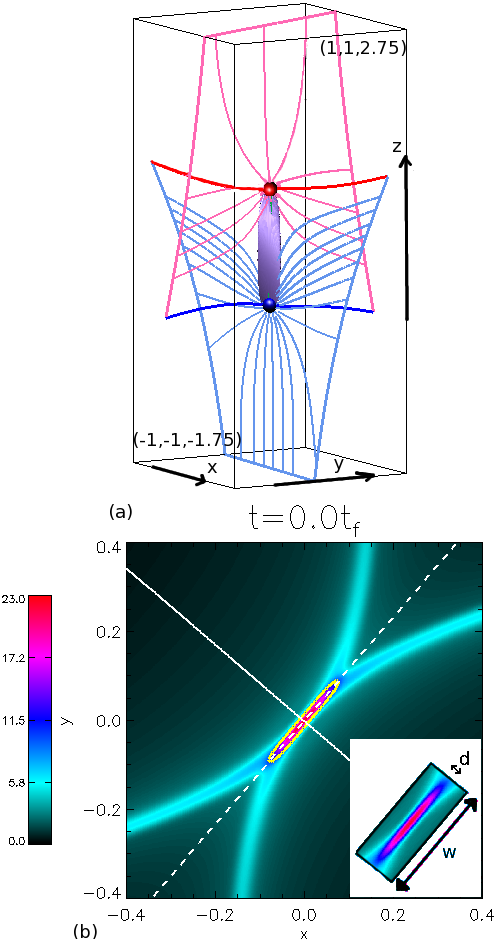}}
\vspace{0.01\textwidth}
\caption{(a) Skeleton of the MHS equilibrium magnetic field with purple isosurface of $j_{\parallel}=10.0$. Also shown are the positive/negative nulls (blue/red spheres) with associated spines (blue/red lines) and separatrix-surface field lines (pale-blue/pink lines) and the separator (green line, hidden by the current layer). The solid pale-blue/pink lines indicate where the separatrix surfaces intersect the boundaries. A movie is available online showing a $360^{\circ}$ view. (b) Perpendicular cut across the MHS equilibrium separator at $z=0.4$ showing contours of $|{\textbf{j}}|$ and white solid/dashed lines going through the depth/across the width of the current layer, respectively. A yellow contour is drawn at $j_{crit}=10$. The inserted image highlights the depth ($d$) and width ($w$) of the current layer in this plane.}\label{fig:equbskel}
\end{figure}

The MHS equilibrium contains two 3D null points located at (0,0,-0.10) and (0,0,1.08) with a separator along the $z$-axis linking them created from the intersection of their separatrix surfaces. (The null points in our model were tracked using the trilinear null finding method of \citet{HaynesP07} and the magnetic skeleton was found using the method described in \citet{Haynes10}). This equilibrium contains a twisted current layer which lies along the separator (Fig.~\ref{fig:equbskel}a, a movie is available online showing a $360^{\circ}$ view of the magnetic skeleton with the current layer). The dominant component of current in the current layer is parallel to the separator. Fig.~\ref{fig:equbskel}b shows a horizontal cut perpendicular to the separator at $z=0.4$ (the location of the peak current along the separator) in which the strong current at the separator (pink contour) and along the separatrix surfaces (cyan curved contours) is visible. In this cut, an insert of the current layer around the separator is included to indicate its width, $w$, and depth, $d$. We have also plotted a yellow contour at $j_{crit}=10$: the current above which the diffusivity is non-zero (see Sect.~\ref{sec:nummodel}). 

In the MHS equilibrium, the separatrix surfaces that create the separator are twisted about it forming cusp regions in planes perpendicular to the separator. Within these cusps the plasma pressure is enhanced and outwith them the plasma pressure is diminished. The equilibrium is not perfect: an infinite time would be required to achieve such a thing \citep{Stevenson15}. Instead, very small localised residual pressure and Lorentz forces remain on the edge of the current layer about the separator and the separatrix surfaces near the separator. Outside the current layer the current is small and the field is in force balance.

\section{Numerical Model}\label{sec:nummodel}
The MHS equilibrium current layer has
length $l_{sep}=1.18$, depth $d=0.06$ and width $w=0.24$ (the latter two dimensions are calculated at $z=0.4$). To study the reconnection that can occur at this separator current layer due to micro-instabilities (modelled by introducing an anomalous diffusivity), we employ the 3D resistive MHD code Lare3d \citep{Arber01}. 

Lare3d solves the MHD equations in a frame that moves with the fluid and then maps the resulting Lagrangian grid geometrically back onto the original Eulerian grid. It uses a staggered grid in which the pressure ($p$), internal energy per unit mass ($\epsilon$) and the density ($\rho$) are defined at the cell centres. The magnetic field components ($B_x, B_y, B_z$) are defined on the cell faces with the \citet{Evans88} constrained transport method for magnetic flux employed to help maintain the solenoidal constraint, $\nabla \cdot {\textbf{B}}=0$. The velocity components ($v_x, v_y, v_z$) are defined on the cell vertices to prevent the checkerboard instability. The following normalised quantities (identified by the hats) represent dimensionless variables used by the code, 
\begin{equation}
{\bf {x}} = L_{n}\hat{\bf {x}}, \qquad {\bf {B}} = B_{n}\hat{\bf {B}} \qquad\text{and}\qquad \rho = \rho_{n}\hat{\rho},
\end{equation}
where ${\textbf{x}}=(x,y,z)$ is the position vector with $L_n$, $B_n$ and $\rho_n$ representing the normalised length, magnetic field and density, respectively. Using these normalising factors, the normalising constants for the velocity, pressure, current, internal energy per unit mass and plasma beta may be written, respectively, as
\begin{equation*}
v_{n} = \frac{B_{n}}{\sqrt{\mu_{0}\rho_{n}}}, \hspace{0.2cm}  p_{n} = \frac{B_{n}^2}{\mu_{0}}, \hspace{0.2cm} j_{n} = \frac{B_{n}}{\mu_{0}L_{n}}, 
\end{equation*}
\begin{equation}
\epsilon_{n} = v_{n}^2 = \frac{B_{n}^2}{\mu_{0}\rho_{n}} \hspace{0.2cm} \text{and}\hspace{0.2cm} \beta = \frac{2\hat{p}}{\hat{B}^2}.
\end{equation}
Note, in the code's dimensionless units the magnetic permeability $\mu_0$=1. This leads to the following resistive normalised MHD equations which are used in Lare3d (note, the hats have been dropped for simplicity)
\begin{equation}
\frac{\rm{D}\rho}{\rm{D}t} = -\rho \nabla \cdot {\bf{v}},
\end{equation}
\begin{equation}
\frac{\rm{D}{\bf{v}}}{\rm{D}t} = \frac{1}{\rho}(\nabla \times {\bf{B}}) \times {\bf{B}} - \frac{1}{\rho} \nabla p + \frac{1}{\rho}{\bf{F}}_{\nu},
\end{equation}
\begin{equation}
\frac{\rm{D}{\bf{B}}}{\rm{D}t} = ({\bf{B}} \cdot \nabla){\bf{v}} - {\bf{B}}(\nabla \cdot {\bf{v}}) - \nabla \times(\eta \nabla \times {\bf{B}}),
\end{equation}
\begin{equation}
\frac{\rm{D} \epsilon}{\rm{D}t} = -\frac{p}{\rho}\nabla \cdot {\bf{v}} + \frac{1}{\rho}H_{\nu} + \frac{j^2}{\rho \sigma},
\label{eq:inte}
\end{equation}
where $t$ is time, ${\textbf{F}}_{\nu} = \nu(\nabla^2{\textbf{v}} + \tfrac{1}{3}\nabla(\nabla \cdot {\textbf{v}}))$ is the viscous force (where $\nu$ is the background viscosity), $\eta=1/(\mu_0 \sigma)$ is the magnetic diffusivity where $\sigma$ is the electric conductivity, $H_{\nu} = \nu(\tfrac{1}{2}e_{ij}e_{ij}-\tfrac{2}{3}(\nabla \cdot {\textbf{v}})^2)$ is the viscous heating term (where $e_{ij}$, the rate of strain tensor, equals $(\partial v_i/\partial x_j) + (\partial v_j/\partial x_i)$) and $\sigma$ is the electric conductivity. The term $j^2/\sigma$ represents Ohmic dissipation and our closure equation is $\epsilon=p/\rho(\gamma - 1)$ where $\gamma$, the ratio of specific heats, equals $5/3$. The value of the background viscosity is $\nu=0.01$. The background viscosity (which is equal to $\rho \nu_k$ where $\nu_k$ is the kinematic viscosity) plays a role in the viscous force, $F_{\nu}$, and viscous heating terms, $H_{\nu}$, discussed previously.

All times in this paper are normalised with respect to $t_f$, the time it would take a fast magnetoacoustic wave to travel from the lower null to the upper null along the path of the separator ($z$-axis) in the initial MHS equilibrium:
\begin{equation}
t_f = \int_{z_l}^{z_u} \frac{1}{\sqrt{c_s(z)^2+c_A^2(z)}}dz = 0.88,
\end{equation}
where $(0,0,z_l)$ and $(0,0,z_u)$ are the equilibrium positions of the lower and upper nulls respectively and $c_s(z)$ and $c_A(z)$ are the sound speed and the Alfv\'en speed at $(0,0,z)$, respectively.

We employ line-tied boundary conditions to prevent energy leaving or entering the domain. Hence, the derivatives of the internal energy per unit mass and the density and all the components of the magnetic field normal to the boundaries are set to zero. The velocity is also set to zero on the boundaries (${\textbf{v}}={\textbf{0}}$).

To gain reconnection, resistive terms are included in the governing equations. We choose to use a non-uniform diffusivity which is zero unless the current is greater than a set amount, $j_{crit}$,
\[ \eta = \left\{
  \begin{array}{l l}
    0 & \quad |{\bf{j}}| < j_{crit},\\
    \eta_d & \quad |{\bf{j}}| \ge j_{crit}.
  \end{array} \right.\]
In our main experiment $j_{crit}=10$ such that diffusion only occurs at the separator current layer (and not on the current on the separatrix surfaces), $\eta_d= 0.001$, corresponding to an average $R_m$ of $10^3$ along the separator, and $\nu=0.01$. An analysis of how varying $\eta_d$, $j_{crit}$ and $\nu$ effect the reconnection rate and the energy conversion is presented in Sect.~\ref{sec:etavisc}.  The behaviour in all our experiments is very similar, so we mainly concentrate on one (the main) experiment in order to describe the basic behaviour that is seen in all experiments. 

The initial setup for all experiments discussed in this paper is a MHS equilibrium. When we apply a non-zero diffusivity, reconnection occurs immediately (as will be discussed in detail in Sect.~\ref{sec:energetics}). If $\eta_d=0$, then the current layer in the equilibrium does not decay \citep[see][]{Stevenson15}; indicating that the numerical diffusion is smaller than our range of values for $\eta_d$.

In the following sections, we describe the basic energetic behaviour and partitioning found in the main experiment (Sect.~\ref{sec:energetics}). Then the nature of the reconnection and magnetic field evolution are studied in Sect.~\ref{sec:phase1}, while Sect.~\ref{sec:etavisc} studies the effects of varying the diffusivity, the size of the diffusion region and the background viscosity on the energetics and reconnection rate. Finally, in Sect.~\ref{sec:conclusions}, the conclusions are presented.

\section{Energetics}\label{sec:energetics}
\begin{figure}[h!]
\centerline{
\includegraphics[width=0.45\textwidth,clip]{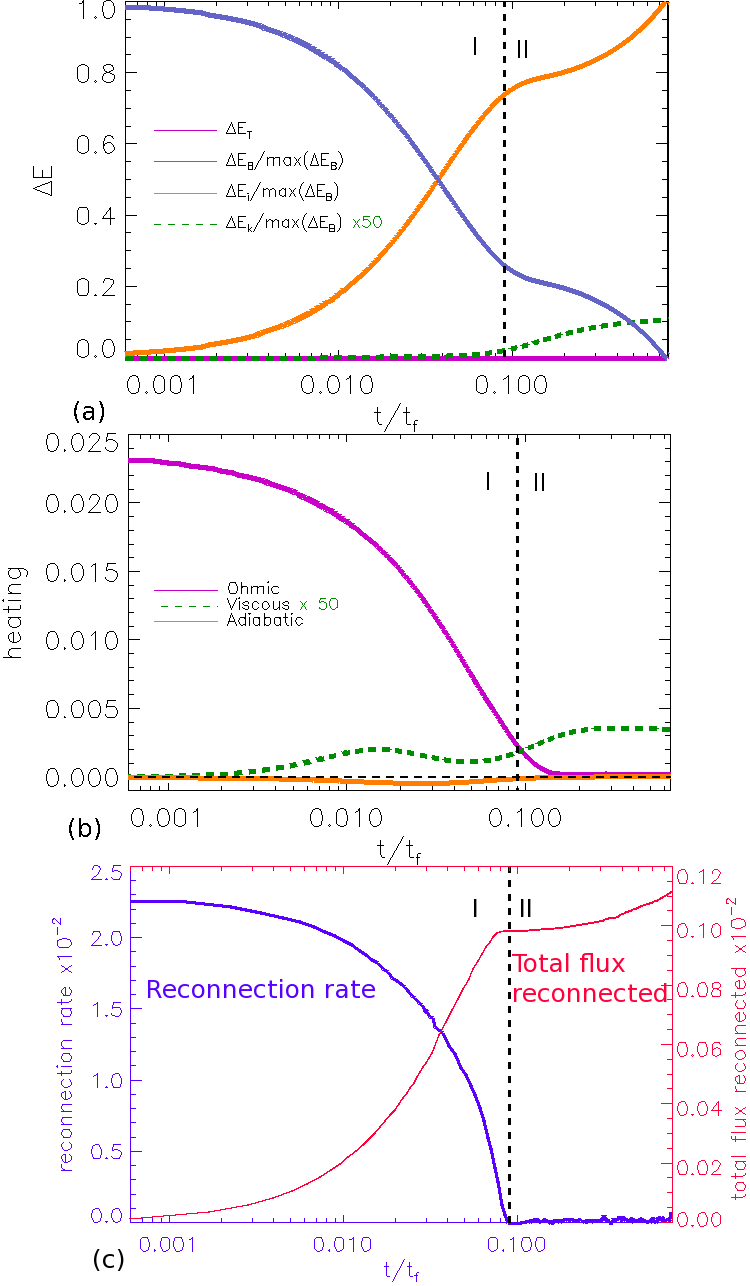}}
\caption{Plots of (a) the change in energies, (b) the instantaneous heating terms and (c) the reconnection rate (blue line) and total flux reconnected (pink line). Note, the kinetic energy and viscous heating curves have been multiplied by fifty and both curves in (c) have been multiplied by one hundred. The black dashed vertical lines highlight where the first phase ends (along with the symbols I and II) and the dashed horizontal line in (b) indicates where zero lies.}\label{fig:energy+rec+heat}
\end{figure}

In order to determine the response of the MHS equilibrium to the introduction of an anomalous diffusivity we consider the system's energetics. The change in energies against time for the main resistive 3D MHD experiment with $\eta_d=0.001$, $j_{crit}=10$ and $\nu=0.01$ are plotted in Fig.~\ref{fig:energy+rec+heat}a, where the magnetic, internal and kinetic energies are normalised to the maximum change in magnetic energy and the normalised kinetic energy is multiplied by fifty for representational purposes. Note, time is plotted on a logarithmic scale.

Since there is no Poynting Flux on the boundaries (i.e., no gain or loss of energy through the boundaries) the change in total energy ($\Delta E_T$) should be zero throughout the experiment. We find that it is zero to within a relative error of $2.1 \times 10^{-5}$\%, indicating that the code properly conserves energy. Fig.~\ref{fig:energy+rec+heat}a shows that in the first $0.09t_f$ of the experiment, magnetic energy ($\Delta E_B$) is converted directly into internal energy ($\Delta E_i$) with only a little going into kinetic energy ($\Delta E_k$). During this time three quarters of the total loss in magnetic energy occurs. As seen from Fig.~\ref{fig:energy+rec+heat}b, this magnetic energy is converted into internal energy via Ohmic dissipation with only 0.23\% going into kinetic energy. 

Fig.~\ref{fig:energy+rec+heat}c shows the reconnection rate and the total flux reconnected during the experiment (calculated from $\int_lE_{\parallel}dl$ along the separator as in \cite{Parnell10a}, an approach that is validated by Fig.~\ref{fig:contepar}). Clearly, the rate of reconnection before $t=0.09t_f$ is dramatically different to that after and so we define this time as the change between two different phases: phase I, the fast-reconnection phase (88\% of the final total reconnected flux is reconnected here) and phase II, the slow impulsive-bursty reconnection phase. 
Negligible reconnection occurs in the first few moments of the second phase (c.f. the flat gradient just after $t=0.09t_f$), but then small amounts of reconnection occur in bursts throughout the second phase. The vertical black dashed line on all these plots denotes where the change of phase occurs. 

The existence of two distinct reconnection phases is not unexpected since the 2D spontaneous reconnection experiments of \citet{FFP12} and \citet{FFP12b} found the same behaviour. \citet{FFP12} (high-beta) found that the fast-reconnection phase was followed by a steady, as opposed to the bursty, reconnection phase we find. In their low-beta model a bursty reconnection phase was found, suggesting that one of the key differences between low and high beta spontaneous reconnection would be a much stronger implusive-bursty reconnection phase after the initial fast-reconnection phase.

Differences between the energetics of the two phases are evident in all graphs in Fig.~\ref{fig:energy+rec+heat}. During phase I, in addition to the reconnection, which begins at a high rate, but decreases rapidly so it is small by the end of the first phase, there is a minimal contribution from the viscous heating term (whose peak during this phase is just $(1/500)^{th}$ of the peak Ohmic heating rate). A small amount of adiabatic cooling is also observed (Fig.~\ref{fig:energy+rec+heat}b) indicating that locally a rapid expansion has occurred within the system. 

As the first phase (in which there is fast Ohmic dissipation) comes to an end at $t=0.09t_f$, the gradients of the magnetic and internal energies change and the Ohmic heating rate decreases: most of the current above $j_{crit}$ in the separator current layer has been dissipated. In phase II, the kinetic energy, which until now has been slowly building, increases sharply associated with the presence of flows in the system following the rapid adiabatic cooling (see \citet{Stevenson15_jgrb} for further details of the waves and flows created by this reconnection). This increase in kinetic energy is reflected by an increase in viscous heating to $8 \times 10^{-5}$, however, the amount of Ohmic heating is always greater than the viscous heating throughout both phases due to the high value of the plasma beta (the mean value of the plasma beta is $\bar{\beta}=4.8$ at $t=0t_f$) which makes it hard for large waves to be produced. In Fig.~\ref{fig:energy+rec+heat}b), the viscous heating has been multiplied by 50, which is why, in this plot, the viscous heating appears to be greater than the Ohmic heating in phase II. The kinetic energy, and viscous heating terms, level out towards the end of the experiment indicating that a phase of slow reconnection has started.

The reconnection in this experiment, therefore, occurs in two phases: the first phase is a highly dynamic fast-reconnection phase with low velocities which is dominated by Ohmic heating ($0t_f \le t \le 0.09t_f$) and the second phase is a slow impulsive-bursty reconnection phase which has a comparatively low level of Ohmic heating, but with higher velocities than in phase I providing greater amounts of viscous heating ($0.09t_f < t \le 0.76t_f$ where $t=0.76t_f$ is where we stop the experiment since at this time the waves, launched by the reconnection, approach the boundaries of the domain, see \citet{Stevenson15_jgrb} for more details). 

In this paper, we focus on the properties of the reconnection in both phases of the experiments. In a follow up paper we will detail the properties of the waves which travel out from the diffusion site, due to the sudden lack of force balance, and set up flows in the system.

\section{Nature of the Reconnection}\label{sec:phase1}
\subsection{Magnetic Field Evolution}\label{sec:phase1-magfield}
\begin{figure}[h!]
\centerline{
\includegraphics[width=0.35\linewidth,clip]{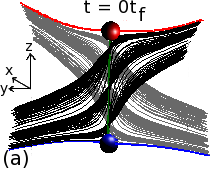}
\includegraphics[width=0.35\linewidth,clip]{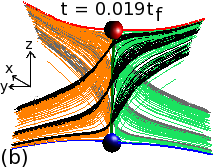}}
\vspace{0.5cm}
\centerline{
\includegraphics[width=0.35\linewidth,clip]{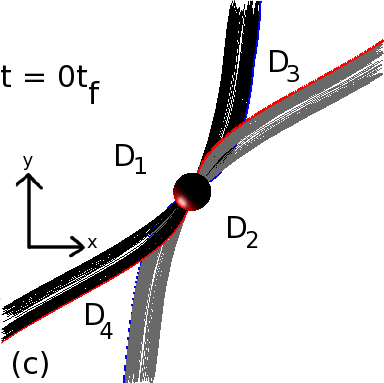}
\includegraphics[width=0.35\linewidth,clip]{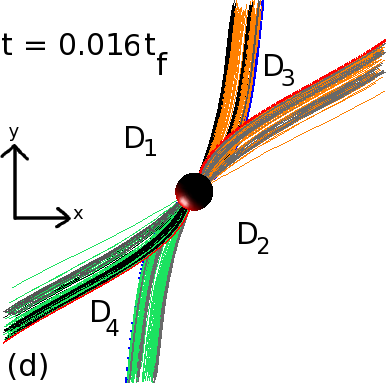}}
\vspace{0.5cm}
\centerline{
\includegraphics[width=0.35\linewidth,clip]{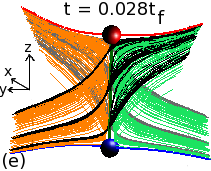}
\includegraphics[width=0.35\linewidth,clip]{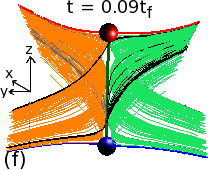}}
\vspace{-.25cm}
\caption{Positive/negative nulls (blue/red spheres) with spines (blue/red lines) and dark green separator. Black and grey field lines are drawn, at $t=0t_f$, within the regions which lie outwith the cusps (i.e., $D_1$ and $D_2$, respectively). When the grey/black field lines reconnect their colours change to orange/green and they lie in regions $D_3$/$D_4$, respectively. (a) and (c) $t=0t_f$ (side and top views), (b) and (d) $t=0.019t_f$ (side and top views), (e) $t=0.028t_f$ and (f) $t=0.09t_f$.}\label{fig:recskel}
\end{figure}

Reconnection at a 3D magnetic separator is found to have several similar characteristics to 2D null-point reconnection (such as the flux from one pair of oppositely-situated flux domains is transferred into another pair of oppositely-situated flux domains), shown in Fig.~\ref{fig:recskel}. Fig.~\ref{fig:recskel} displays the nulls, their spines, the separator and the same sample field lines (drawn initially in the two flux domains outwith the cusp regions, i.e., $D_1$ and $D_2$, respectively) at four times throughout the fast-reconnection phase, which lasts up until $t=0.09t_f$. The earliest two times are plotted from both side and top views to show which domains the field lines lie in before and after the reconnection occurs. In these plots, field lines which have not reconnected are drawn as black or grey lines, if they lie in domains $D_1$ and $D_2$, respectively. Once reconnected, these field lines are coloured orange and grey and lie in domains $D_3$ and $D_4$, respectively. This reconnection is not undertaken by a pairwise matching of field lines, but by continual reconnection of field lines during their passage through the diffusion region. 

As shown in Fig.~\ref{fig:energy+rec+heat}, and confirmed in Fig.~\ref{fig:recskel}, much of the reconnection occurs within the first phase. At $t=0.09t_f$, there are, however, some field lines which have not yet reconnected (e.g., the remaining black and grey field lines in Fig.~\ref{fig:recskel}f). These are an example of the field lines that are reconnected during the second phase of the reconnection. Movies, available online, show the evolution of all of these field lines from side and top views throughout the reconnection experiment.

One important point to note is that the black and grey (pre-reconnection) field lines do not lie along the separator, but cross by it, entering almost parallel to the spine of the lower null and then leaving parallel to the upper null's spine. This means that the field lines twist through $\approx 180^\circ$ about the separator before they reconnect. After they reconnect (orange and green lines) the field lines still run almost parallel to the lower null's spine, but now leave running parallel to the other side of the upper null's spine. To do this the field lines run almost straight up the separator and barely twist around it at all. 

\begin{figure}[h!]
\centerline{
\includegraphics[width=0.4\textwidth,clip]{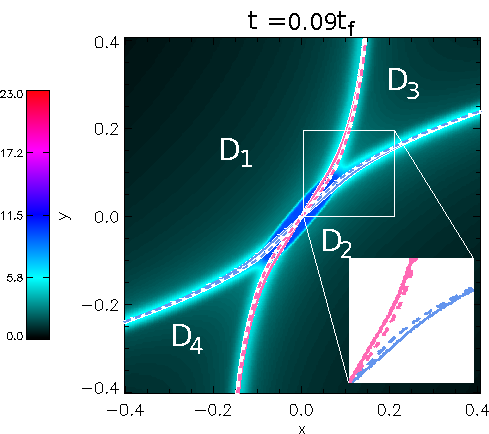}}
\caption{Contours of $|{\textbf{j}}|$, in a cut at $z=0.4$ across the separator, at $t=0.09t_f$. The intersections of the lower (pale-blue) and upper (pink) null's separatrix surfaces with this plane are also plotted at $t=0t_f$ (dashed lines) and at $t=0.09t_f$ (solid lines) on top of white lines so they are distinct from the contours. The separatrix surfaces split the cut up into four regions marked as $D_1$, $D_2$, $D_3$ and $D_4$. The insert shows the separatrix surfaces of the equilibrium field and the field at $t=0.09t_f$, in the region $0.0 \le x,y \le 0.2$, close up.}\label{fig:jphase1}
\end{figure}

The reconnection at the separator rapidly dissipates the current above $|{\bf{j}}|=j_{crit}$ in the MHS equilibrium current layer, which lies along the length of the separator. The current at $t=0.09t_f$, in a cut across the separator at $z=0.4$ (Fig.~\ref{fig:jphase1}, c.f. with Fig.~\ref{fig:equbskel}b which is the same plot, but in the initial equilibrium state) decreases at the separator current layer, but does not decrease along the separatrix surfaces since the value of current here is less than $j_{crit}$. The current along the entire length of the separator diminishes quickly from its peak of 22.4 and mean of 19.5 during the first phase until it is just below the value of $j_{crit}$ at about $|{\textbf{j}}|=9.8$ all the way along the separator. This is an indication that separator reconnection occurs along nearly the whole length of the separator and not just at one point as in null-point reconnection (in agreement with \citet{Parnell10a}). After the first phase, the current does not really change along the separator. 

The dissipation of the current at the separator causes the separatrix surfaces of the nulls to open up slightly in the cusp regions, as flux is transferred from domains $D_1$ and $D_2$ outside the cusps to the domains $D_3$ and $D_4$ that are inside the cusps. This behaviour is visible in Fig.~\ref{fig:jphase1} where the intersections of the separatrix surfaces of both nulls with this plane are plotted at $t=0t_f$ (dashed pale-blue/pink lines) and at the same time that the contours of $|{\bf{j}}|$ are drawn, $t=0.09t_f$ (solid pale-blue/pink lines). The separatrix surfaces of the nulls are shown to open up slightly within the cusp regions (for closeup see Fig.~\ref{fig:jphase1} insert), as a consequence of the reconnection. 

The reconnection causes the nulls to move slightly apart in the $z$-direction, with the maximum length of the separator throughout the experiment just $1.005$ times greater than the equilibrium separator length. The lower null moves downwards from its initial position, along the $z$-axis, by $0.1L_0$ in the first $0.03t_f$, whereas the upper null moves upwards $0.08L_0$ along the $z$-axis, in the first $0.01t_f$. After these times both nulls move very slowly away from each other (the lower null moves a further distance of $1\times 10^{-4}L_0$ and the upper null moves a further $5\times 10^{-5}L_0$) during the rest of the experiment. 

\subsection{Nature of $E_{\parallel}$}\label{sec:phase1-epar}
\begin{figure}[h!]
\centerline{
\includegraphics[width=0.4\textwidth,clip]{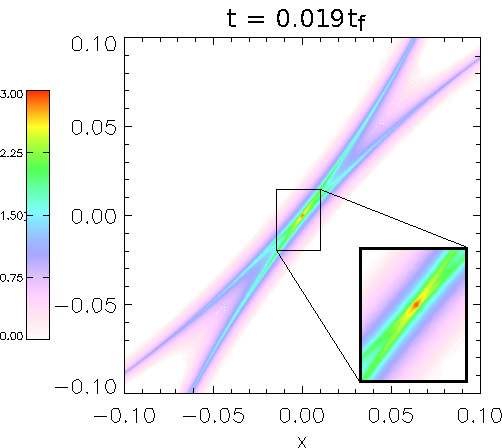}}
\caption{Contours of the integral of $E_{\parallel}$ along field lines which thread this plane perpendicular to the separator at $z=0.4$, at $t=0.019t_f$. The insert shows a close up view of the region highlighted by the box around the separator.}\label{fig:contepar}
\end{figure}

In 3D, a non-zero integral of $E_{\parallel}$ ($=\eta j_{\parallel}$) along a field line indicates the existence of reconnection \citep{Schindler88,Hornig96}. Furthermore, in a situation where there is a single simple diffusion region, such as we have here, the maximum integral of $E_{\parallel}$ identifies the main reconnection site, as well as the reconnection rate \citep{Hesse93}. Since all the field lines that thread the non-ideal region ($\eta_d \ne 0$) thread planes that cross perpendicular to the separator, we determine the integral of $E_{\parallel}$ along field lines that thread the plane $z=0.4$ at $t=0.019t_f$ during phase I. A contour plot of $\int_lE_{\parallel}dl$ (Fig.~\ref{fig:contepar}) shows that the strongest reconnection occurs at the separator (in agreement with \citet{Parnell10a}). The insert in this figure shows a close up of the contours around the separator in this cut and highlights the strength of the localised peak of the integral of $E_{\parallel}$ along field lines at the separator. Weaker reconnection occurs on neighbouring field lines creating non-zero $\int_lE_{\parallel}dl$ regions along the separatrix surfaces that form the separator. Now that we have confirmed that the reconnection in our experiment is separator reconnection, we look at the time evolution of various parameters along the separator to determine more details about the nature of the reconnection itself.

Fig.~\ref{fig:timeslice}a shows how the parallel component of the electric field ($E_{\parallel}$) along the separator evolves in time as the reconnection proceeds. The separator is normalised such that it has length one for all time according to the equation $z^* = (z-z_{l})/l_{sep}$ where $z_{l}$ (the $z$-coordinate of the lower null) and $l_{sep}$ (the length of the separator) take their respective values for the frame being considered. 
The strongest reconnection occurs between about $z^*=0.16$ and $z^*=0.7$. Midway between these points is close to where the initial peak in current lies indicating that the peak reconnection is occurring where the current is strongest, as expected. Note that the strongest values of $E_{\parallel}$ lie midway along the separator away from the nulls indicating that the null points are not involved in separator reconnection \citep[in agreement with][]{Parnell10a}. 

After about $t=0.03t_f$, the peak $E_{\parallel}$ (Fig.~\ref{fig:timeslice}a) is about half of what it was at the start of the experiment and decreases over the next $t=0.06t_f$ to almost nothing. During the second phase, patches of $E_{\parallel} > 0$ are visible in Fig.~\ref{fig:timeslice}a. These regions, which have $|{\bf{j}}| \ge j_{crit}$, exist very briefly since the current in excess of $j_{crit}$ is immediately dissipated due to the nature of the non-uniform diffusivity. Although these small reconnection events do not exist for long (compared to the reconnection in phase I) there are a significant number of these events such that the total flux reconnected slowly grows in phase II, as shown in Fig.~\ref{fig:energy+rec+heat}c. 

The solid black curve in Fig.~\ref{fig:timeslice}a is the zeroth contour of the discriminant of the perpendicular component of the magnetic field (${\bf{B}}_{\perp}$), along the separator. This contour highlights that the projected magnetic field lines are locally elliptic (O-type), in planes perpendicular to the separator, in the regions where $E_{\parallel}$ is strongest, during phase I. Elsewhere in phase I, and everywhere in phase II, the projected magnetic field lines are locally hyperbolic (X-type) in planes perpendicular to the separator. This behaviour is consistent with that discussed in Sec~\ref{sec:phase1-magfield} where the pre-reconnected field wound around the separator $\approx 180^\circ$, whilst after reconnection the field lines ran more parallel to the separator and did not cross it. This has been seen before in separator reconnection experiments by \citet{Parnell10a} who found that cuts across the separator, where the reconnection was strongest, corresponded to 2D O-type magnetic field configurations. 

\subsection{Behaviour of the Local Reconnection Flows}\label{sec:phase1-local-flows}
\begin{figure*}[h!]
\centerline{
\includegraphics[width=0.5\textwidth,clip]{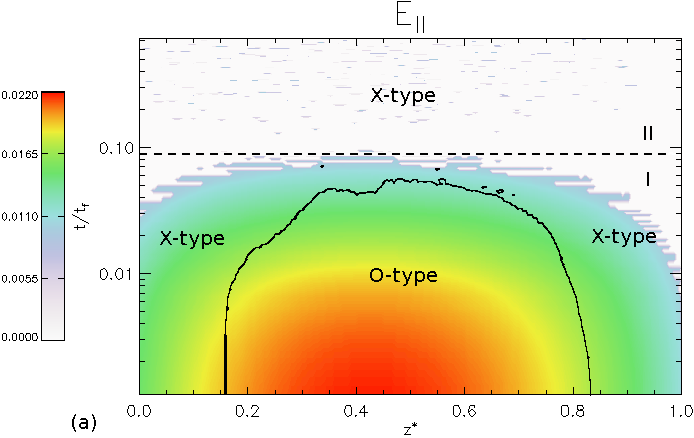}
\includegraphics[width=0.5\textwidth,clip]{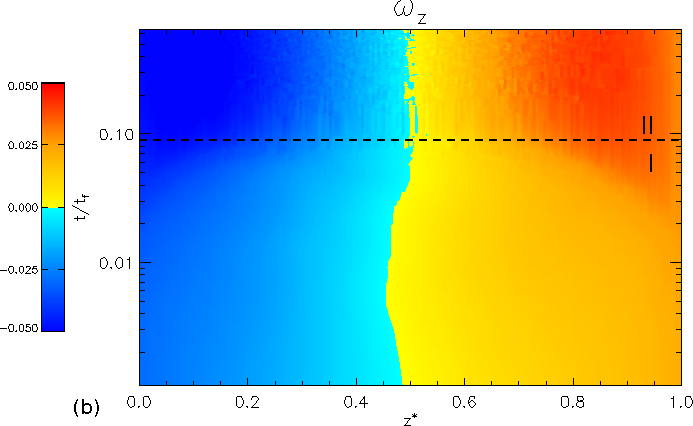}}
\centerline{
\includegraphics[width=0.5\textwidth,clip]{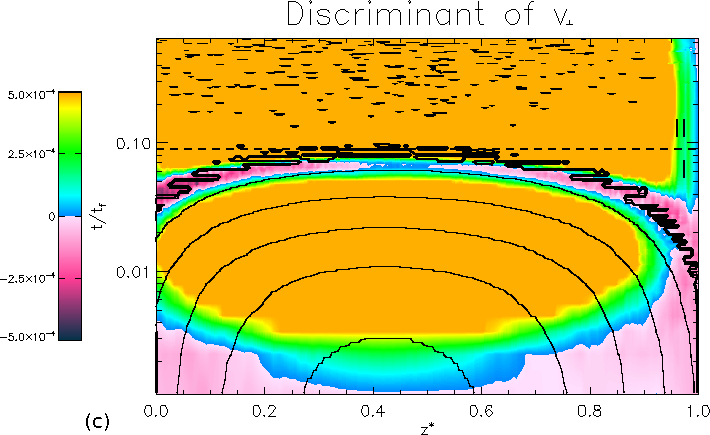}
\includegraphics[width=0.5\textwidth,clip]{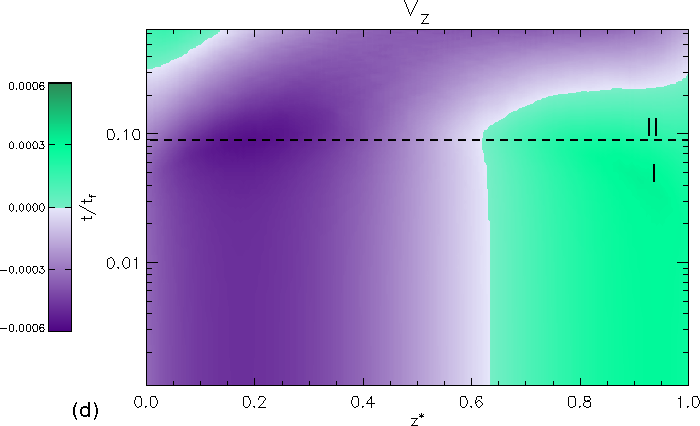}}
\caption{The time evolution of (a) $E_{\parallel}$ (coloured contours) with the zeroth contour of the discriminant of ${\bf{B}}_{\perp}$ over plotted as black lines and the nature of the 2D magnetic field lines in planes perpendicular to the separator annotated on the plot, (b) the component of the vorticity parallel to the separator ($\omega_z=(\nabla \times {\bf{v}})_z$), (c) the discriminant of ${\bf{v}}_{\perp}$ and contours of $E_{\parallel}$ (shown in Fig.~\ref{fig:timeslice}a) are over plotted as black lines and (d) the $z$-component of the velocity, $v_z$, along the separator (which is normalised to lie between $z^*=0$ and $z^*=1$). The vertical axis on each plot is logarithmic to highlight the behaviour during the first phase and the dashed line (and symbols I and II) indicate where the first phase ends and the second phase begins.}\label{fig:timeslice}
\end{figure*}

Unlike in 2D reconnection, the flows local to separator reconnection are not simple stagnation flows.
\citet{Schindler88,Hesse88,Hornig96} have shown that, in 3D, the flow in the diffusion region counter-rotates such that a continuum of field lines discontinuously reconnect at the separator in a manner that is not pairwise. This is indeed the case for separator reconnection, as evidenced by the vorticity parallel to the separator, $\omega_z = (\nabla \times {\bf{v}})_z$, (Fig.~\ref{fig:timeslice}b). Therefore, in 3D separator reconnection the magnetic field lines undergo a rotational slippage about the separator, which is not surprising since they are effectively undoing their $\approx 180^\circ$ twist about the separator.

Indeed, Fig.~\ref{fig:timeslice}b shows that the streamlines curl around the separator in opposite directions above and below a point initially around $z^*=0.48$ (midway between where the current and pressure are initially highest on the separator). This point, about which the streamlines counter-rotate, moves downwards along the separator during phase I (when the value of $E_{\parallel}$ here is strong) before returning to around $z^*=0.49$ near the end of phase I where it remains during phase II. Thus, the point on the separator about which the flow counter-rotates is close to, but not coincident with the position of the peak reconnection along the separator.

To discover more about the nature of the flow we also look at the time evolution of the discriminant of the component of velocity perpendicular to the separator (${\bf{v}}_{\perp}$) (coloured contours in Fig.~\ref{fig:timeslice}c). This plot indicates that, in planes perpendicular to the separator, the flow starts off O-type locally about the separator, although the velocities here are very small and so there is effectively no flow. The flow perpendicular to the separator then becomes X-type away from the nulls during the first phase where strong reconnection is occurring (the black lines in Fig.~\ref{fig:timeslice}c are contours of $E_{\parallel}$ along the separator, as determined in Fig.~\ref{fig:timeslice}a). As the reconnection rate during phase I decreases the perpendicular flow once again becomes O-type briefly. For the second phase, the local perpendicular flow returns to X-type about the separator along almost its entire length except for very close to the upper null.  

\citet{Parnell10a} studied the local velocity flow and local magnetic field structure in planes perpendicular to a 3D separator during separator reconnection and found a clear-cut relationship between velocity flow and magnetic field local to the separator: when the local flow is O-type the local magnetic field is X-type, and vice versa. Here, comparing the contours of the discriminant of ${\bf{v}}_{\perp}$ (Fig.~\ref{fig:timeslice}c) with the contours of the discriminant of ${\bf{B}}_{\perp}$, black lines on Fig.~\ref{fig:timeslice}a, we see that there is no clear correspondence between the nature of the local flow and field about the separator. At the start of the experiment, the magnetic field local to the separator is strongly O-type (away from the nulls where, by definition, it must be X-type), but the velocities are too small to really claim that the flow is actually O-type. The O-type magnetic field persists on the separator and the velocity becomes X-type along almost the entire length of the separator (in agreement with \citet{Parnell10a}). Briefly, when $E_{\parallel}$ is decreasing rapidly to zero, the flow about the separator becomes O-type: at this time the perpendicular magnetic field is already X-type (again in agreement with \citet{Parnell10a}). After this an X-type velocity flow then returns and persists throughout the weak bursty reconnection phase (phase II) where the perpendicular magnetic field local to the separator is also X-type along its whole length (contrary to \citet{Parnell10a}). Thus, the relationship between the local flow and field within the diffusion region is not as simple as \citet{Parnell10a} state, however, during rapid reconnection the configuration does tend to involve field lines with a twist that are untwisted by a counter-rotating stagnation-type flow.

We have considered the behaviour of the flows in to and out from the separator, but what are they like along the separator? (Fig.~\ref{fig:timeslice}d) shows the $z$-component of velocity along the separator, which, for the whole of phase I, is directed outwards towards the nulls from a point at $z^*=0.63$. This location is neither coincident with the location of the peak reconnection, nor with the point of counter-rotation, nor with the peak pressure along the separator (which is located at about $z^*=0.57$). The point of divergence of this flow abruptly changes as phase II starts and gradually moves up to the upper null such that along the separator the flow is almost entirely downwards. Towards the end of phase II, near the lower null an upflow along the separator is formed. 

So except for a brief period during which the reconnection is rapidly decelerating towards the end of phase I, the local flow about the separator is essentially a form of counter-rotating stagnation flow in which the field in the diffusion region is swept into the separator whilst rotating and then outwards down towards the nulls and away from the separator, again whilst rotating. Due to the lack of co-location of the counter-rotation point and the outflow point of flow along the separator, in the region of peak reconnection the local flows are more complicated.  Note, details of the nature of the flows further out from the separator are discussed in \cite{Stevenson15_jgrb}.

Finally, the $z$-component of the curl of ${\textbf{B}}$ along the separator (not shown) decreases over time, revealing that the magnetic field relaxes as the reconnection proceeds. So, the magnetic field behaviour is as expected and is consistent with the field relaxing, via reconnection, from a twisted state. 

\section{Parameter Analysis:  $\eta_d$, $j_{crit}$ and $\nu$}\label{sec:etavisc}
In order to investigate the effects of (i) the anomalous diffusivity $\eta_d$, (ii) the size of the diffusion region (by changing the value of $j_{crit}$) and (iii) the background viscosity $\nu$, we run additional experiments in which each of these parameters is varied in turn. The results from these runs are discussed in the subsections below. 

Experiments with three different values of diffusivity ($\eta_d=0.0005$, $0.001$ - main experiment and $0.002$) with $j_{crit}=10$ and $\nu=0.01$, then three different values of $j_{crit}$ ($j_{crit}=10$ - main experiment, $j_{crit}=7.5$ and $j_{crit}=8.5$) with $\eta=0.001$ and $\nu=0.01$ and finally three different values of $\nu$ are studied ($\nu=0.005$, $0.01$ -main experiment and $0.02$) with $\eta=0.001$ and $j_{crit}=10$ are studied. None of the plasma parameters, nor the initial magnetic field have changed from the original experiment in these runs except for those stated here. 

\subsection{Effects of Varying $\eta_d$, $j_{crit}$ and $\nu$ on the Reconnection Rate}
\begin{figure*}[h!]
\centerline{
\includegraphics[width=0.33\textwidth,clip]{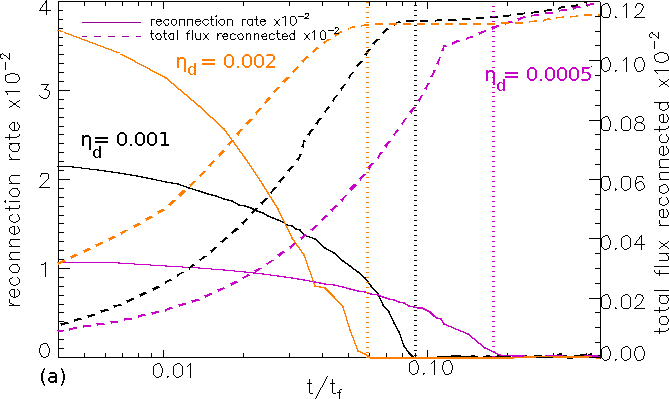}
\includegraphics[width=0.3\textwidth,clip]{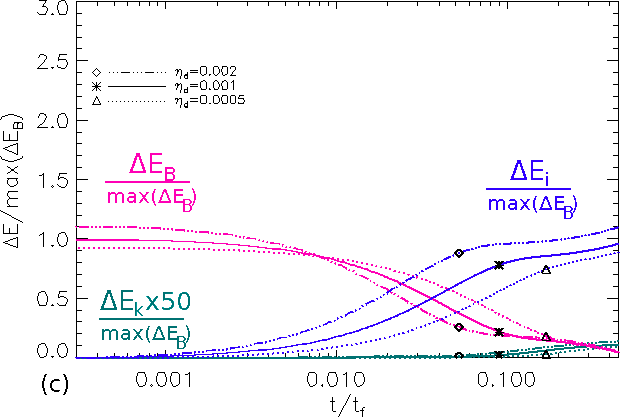}
\includegraphics[width=0.3\textwidth,clip]{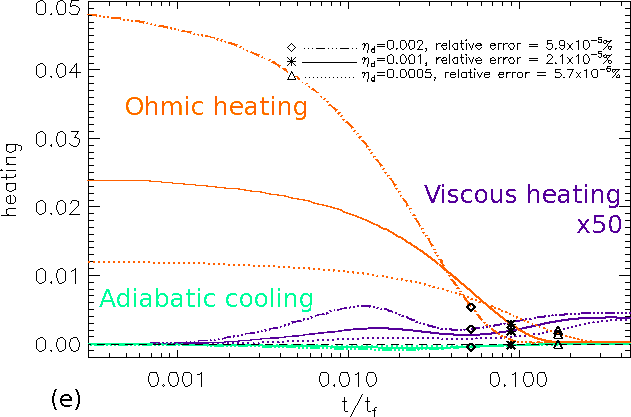}}
\centerline{
\includegraphics[width=0.33\textwidth,clip]{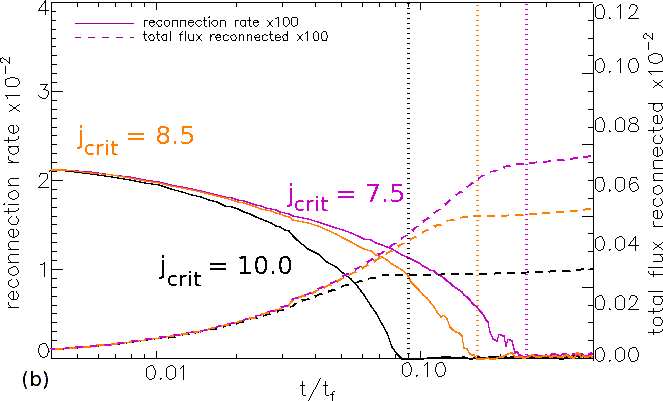}
\includegraphics[width=0.3\textwidth,clip]{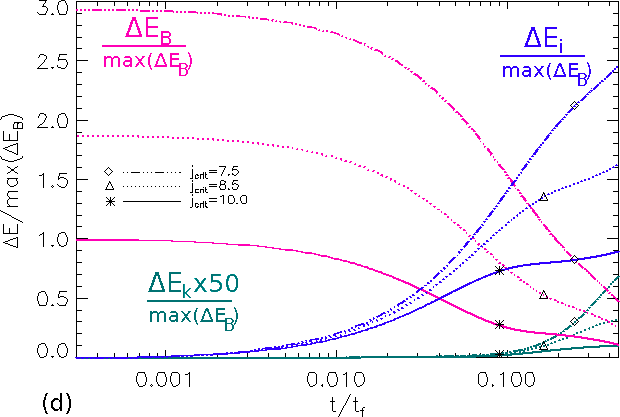}
\includegraphics[width=0.3\textwidth,clip]{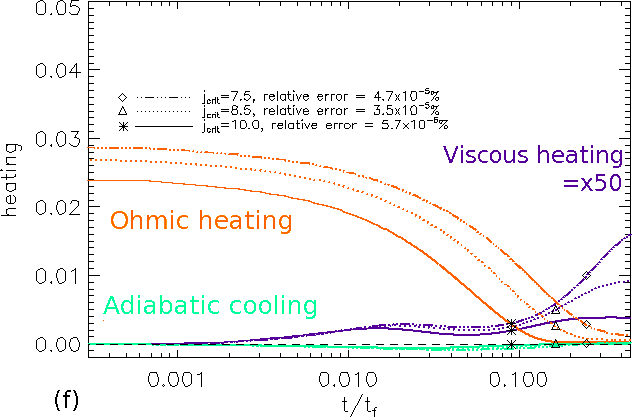}}
\centerline{
\includegraphics[width=0.33\textwidth,clip]{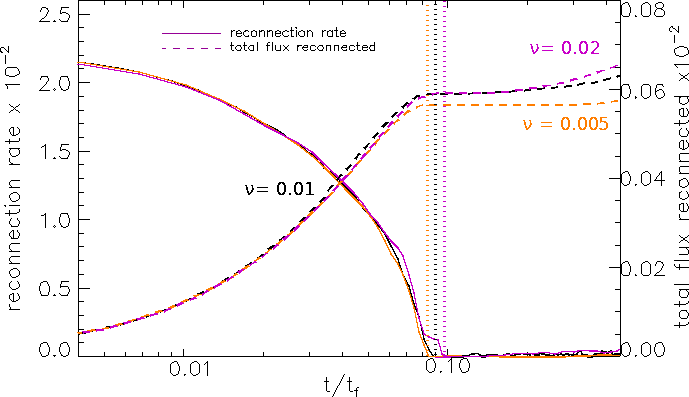}
\includegraphics[width=0.3\textwidth,clip]{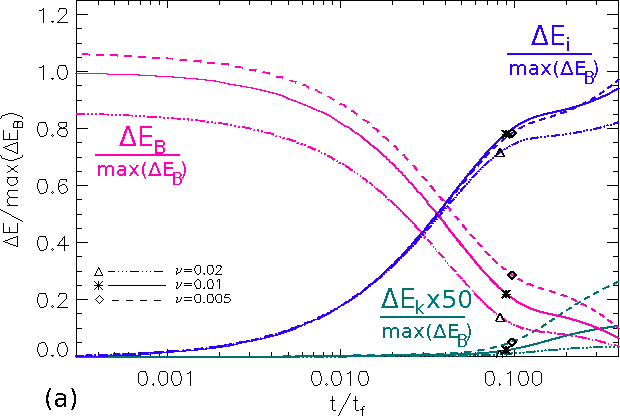}
\includegraphics[width=0.3\textwidth,clip]{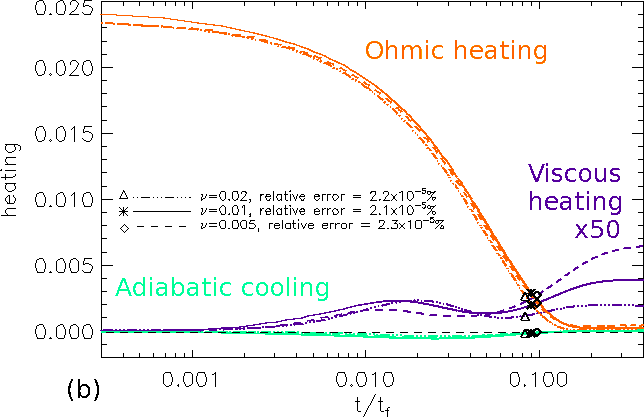}}
\caption{First column: plots of the reconnection rate (solid lines) and the total flux reconnected (dashed lines), multiplied by 100, second column: the energy and third column: the heating terms all plotted against time for experiments with (top row)  $\eta_d=0.0005$, $\eta_d=0.001$ and $\eta_d=0.002$ with $j_{crit}=10$ and $\nu=0.01$, (middle row) $j_{crit}=7.5$, $j_{crit}=8.5$ and $j_{crit}=10$ with $\eta_d=0.001$ and $\nu=0.01$, (bottom row) $\nu=0.005$, $\nu=0.01$ and $\nu=0.02$ with $\eta_d=0.001$ and $j_{crit}=10$. In plots (a) to (c) dashed vertical lines (coloured to match the respective experiment) highlight the time at which phase I ends and phase II begins for each experiment. In plots (d) to (i) symbols (annotated on the plots) indicate where phase I ends and phase II begins.}\label{fig:varyetajcrit}
\end{figure*}

First, we investigate how the reconnection rate and the total flux reconnected depend on $\eta_d$, $j_{crit}$ and $\nu$. As one might expect, the higher the value of $\eta_d$, the faster the reconnection, but the shorter the fast-reconnection phase (phase I) (Fig.~\ref{fig:varyetajcrit}a). The total amount of reconnected flux in all three cases is similar, but appears to marginally increase as $\eta_d$ increases, except in the experiment with $\eta_d=0.002$. We believe this occurs since near the start of the experiment with the highest anomalous diffusivity ($\eta_d=0.002$) additional nulls appear in the system. These nulls, which are formed in opposite-sign pairs, appear close to the locations of the original nulls and lead to the creation of extra intercluster separators in addition to the original (intercluster) separator that linked the two original null points/null point clusters. To calculate the reconnection rate and total flux reconnected, we assume that there is just one reconnection site; the original separator. However, when there are multiple separators/reconnection sites, this assumption breaks down, and so we may not be identifying all the reconnection that occurs in the experiment with $\eta_d=0.002$.  This is the only experiment discussed here which displays this behaviour. The creation of these multiple nulls and separators, and the reconnection associated with them, will be studied in a follow up paper. 

From Fig.~\ref{fig:varyetajcrit}b, we see that the smaller the diffusion region (i.e., the higher the value of $j_{crit}$) the shorter the fast-reconnection phase (phase I). This, of course, is expected as a smaller diffusion region contains less flux to reconnect. However, the peak reconnection rate is unaffected by the value of $j_{crit}$. Thus, the total amount of flux reconnected increases as $j_{crit}$ is lowered.

From Fig.~\ref{fig:varyetajcrit}c, we find that varying the value of $\nu$ has little effect on the reconnection rate during the fast-reconnection phase. During phase II a smaller viscosity is associated with a marginally faster reconnection rate. This is because the resulting flows that drive the steady-state reconnection of phase II are stronger in a fluid that is less viscous. This leads to more flux being reconnected overall in the case with the lowest viscosity.

\subsection{Effects of Varying Diffusivity $\eta_d$ on the Energetics}\label{sec:picketa}
For the three experiments with varying $\eta_d$, the same basic behaviour of energies is found as that seen in the main experiment (Figs.~\ref{fig:varyetajcrit}d and \ref{fig:varyetajcrit}g). The change in energies are normalised to the maximum change in the magnetic energy of the main experiment (where $\eta_d=0.001$, $j_{crit}=10$ and $\nu=0.01$) so that the energy release of all experiments can be compared. As before, most of the magnetic energy lost is converted directly into internal energy with only a little kinetic energy (multiplied by fifty for representational purposes) generated (Fig.~\ref{fig:varyetajcrit}d). Varying $\eta_d$ leads to a change in the rate of loss of magnetic energy: naturally, the experiment with the highest $\eta_d$ experiences the most rapid loss, but the shortest fast-reconnection phase (phase I). 
The total loss of magnetic energy increases as $\eta_d$ increases. 

Most of the additional energy comes directly from Ohmic heating. A small contribution comes from viscous heating (multiplied by fifty for representational purposes), due to some wave damping occurring predominantly during phase II of each experiment (Fig.~\ref{fig:varyetajcrit}g). The rate of viscous heating is very similar in all three experiments suggesting that the more rapid reconnection has not led to the creation of larger perturbations. In each experiment, a small amount of adiabatic cooling occurs in the later part of phase I. This occurs due to the sudden expansion of the field about the separator current layer. 

From Fig.~\ref{fig:varyetajcrit}g, it appears that the peak Ohmic heating rate is basically linearly proportional to $\eta_d$, as is the duration of phase I. However, the duration of the main Ohmic heating period (which, as shown in Sect.~\ref{sec:energetics}, is slightly longer than phase I) does not vary in the same way and is shorter for large $\eta_d$ than expected from a linear fall off. The relative error in the total energy is small and of the same order for all three experiments.

\subsection{Effects of Varying $j_{crit}$ on the Energetics}
Here, we consider the experiments with varying $j_{crit}$. 
From Fig.~\ref{fig:varyetajcrit}e, we see that having a lower value of $j_{crit}$ means that the total loss in magnetic energy increases and it takes longer for the majority of the magnetic energy to be converted into internal and kinetic energy. Furthermore, the kinetic energy is greater as $j_{crit}$ decreases. These differences are not surprising, since a lower $j_{crit}$ creates a larger diffusion region in which more current can be dissipated and creates a greater loss in force balance of the system. We find that the free energy released, (calculated as a percentage of the energy released if a uniform background diffusivity was used), increases as $j_{crit}$ decreases, with $3\%$ of the free energy released when $j_{crit}=10$, $5\%$ when $j_{crit}=8.5$ and $8\%$ when $j_{crit}=7.5$.

In Fig.~\ref{fig:varyetajcrit}h, it is no surprise to see that the initial amount of Ohmic heating increases as $j_{crit}$ decreases. Additionally, a lower initial $j_{crit}$ leads to greater viscous heating associated with the larger kinetic energies in the system.

\subsection{Effect of Varying $\nu$ on the Energetics}
The effects of varying the value of the background viscosity, $\nu$, has little effect on the total loss in magnetic energy (Fig.~\ref{fig:varyetajcrit}f). This is reflected in the Ohmic dissipation (Fig.~\ref{fig:varyetajcrit}i), which appears to be basically the same in all experiments. The proportion of magnetic energy converted to kinetic energy during the experiment does depend on viscosity with a larger $\nu$, corresponding to greater viscous heating in phase I, leading to smaller kinetic energy since the waves are damped to a greater extent. This means that near the end of phase I, and throughout phase II, the kinetic energy is greatest for experiments with lower $\nu$. In phase II, we find that the larger the $\nu$ the weaker the Ohmic heating. The adiabatic cooling term appears to be unaffected by the size of the viscosity.

\section{Conclusions}\label{sec:conclusions}
In this paper, we have studied the properties of spontaneous (undriven) reconnection at a 3D separator current layer using a resistive MHD code. We start from a system containing free energy, which is in MHS equilibrium everywhere save for very small forces at the current enhancements about the separator and separatrix surfaces: a perfect equilibrium would take an infinite time to form \citep{Stevenson15}.  An anomalous diffusivity, $\eta_d$, is used, to mimic the onset of micro-instabilities, which only acts where the current is greater than a set amount, $j_{crit}$, such that reconnection only occurs at the separator current layer and, for low enough $j_{crit}$, a little way along the separatrix surfaces. 

In our experiments, unlike driven separator reconnection experiments, such as \cite{Haynes07,Parnell10a}, the reconnection starts immediately and its rate is dictated simply by the value of the anomalous diffusivity and is not influenced by an external driver.  We find that the reconnection occurs in two distinct phases; a short fast-reconnection phase followed by a longer slow impulsive-bursty-reconnection phase. Such a partitioning into distinct phases is the same as that found in 2D spontaneous reconnection experiments at null points \citep[e.g.,][]{FFP12,FFP12b}, but here an impulsive bursty regime is found even at high beta. 

In the main experiment, where $\eta_d=0.001$, $j_{crit}=10$ and $\nu=0.01$, most $(88\%)$ of the reconnection occurs during the first short phase, that lasts just $0.09t_f$, in which the current is rapidly dissipated away from the current layer. During this phase magnetic energy is mainly $(99.77\%)$ converted directly into internal energy via Ohmic dissipation. Only a small amount $(0.23\%)$ of energy is converted from magnetic to kinetic energy during this phase and then transferred into internal energy via viscous damping. Additionally, due to a rapid expansion of the plasma as a result of the sudden reconnection, a small amount of adiabatic cooling is observed during phase I. Even during the slow impulsive-bursty reconnection phase, the limited Ohmic heating dominates over the viscous heating. This is likely to be a consequence of the high-beta plasma within which the separator is embedded since, in such a plasma, as a current layer is formed a strong pressure gradient force rapidly develops balancing the Lorentz force making it harder for strong current layers to form than in the equivalent low-beta system. Consequently at a high-beta separator current layer the perturbation caused by the sudden onset of reconnection will be smaller than in the low-beta case. Furthermore, the combination of a smaller perturbation and denser plasma in the high-beta, as opposed to low-beta, system means that little plasma is accelerated and, hence, a low kinetic energy is found. Further studies are needed to establish under what conditions, if any, a low-beta regime produces greater viscous heating than Ohmic heating.   

Although the kinetic energy is relatively small, this in no way implies that waves and flows are not produced. The rapid loss of equilibrium caused as a result of the reconnection in the separator current layer naturally launches waves that generate flows within the system. These waves and flows are discussed in detail in the second paper of this series \cite{Stevenson15_jgrb}.

The rate of reconnection in the first phase is up to 22 times faster than in the second phase, and the maximum reconnection rate increases as $\eta_d$ gets larger. However, the duration of the first phase shortens as $\eta_d$ increases, such that the total amount of flux reconnected in this phase is independent of $\eta_d$. As $j_{crit}$ increases, the length of the first phase also decreases, but here the total amount of flux reconnected in the first phase increases as $j_{crit}$ decreases. Varying the value of the background viscosity has little effect on both the rate of reconnection and the length of the first phase.

In this paper, the reconnection rates we quote are dimensionless. These rates may be scaled to dimensional values by multiplying by $B_nL_n^2/t_n$, where $B_n$ (T), $L_n$ (m) and $t_n$ (s) are the normalising magnetic field, length and time. For example, numerical modelling of separator reconnection at the dayside magnetopause by \citet{Komar13} and \citet{KomarPhD} found separators of length $L_n=223$ Mm within magnetic fields with strengths $B_n=5$ nT. Applying these values, with a time scale of $t_n=1$ s, we find the peak reconnection rate at the start of our main experiment is $5.6 \times 10^6$ V. This value is over 100 times greater than the observed value found by \citet{Chisham04} who looked at high latitude dayside reconnection with northward IMF. Our values better match with those found by \citet{Pinnock03} (which range from $1.1 \times 10^6$ V to $9 \times 10^6$ V) who found the rate of reconnection along the projection of a separator in the ionosphere.

Similarly, our diffusivity ($\eta_d=0.001$) can be scaled to dimensional values in the dayside magnetopause by multiplying by $L_nv_{A_n}/R_m$, where $v_{An}$ is the Alfv\'en speed scaling factor. \citet{KomarPhD} found Alfv\'en speeds of $380$ kms$^{-1}$, hence, our diffusivity $\eta_d=L_nv_{An}/R_m =8.5 \times 10^{10}$ m$^2$s$^{-1}$ (where $R_m=10^3$ as discussed in Sect.~\ref{sec:nummodel}). This value is comparable with that used in \citet{Komar13} ($\eta_K=6 \times 10^{10}$ m$^2$s$^{-1}$) who studied similar experiments to those detailed in \citet{KomarPhD}.

During phase I, the strength of the reconnection, which occurs asymmetrically along the entire length of the separator with the strongest reconnection occurring away from the null points, decreases and only weak, short-lived reconnection events occur in phase II. During this second phase, the impulsive bursts of reconnection occur randomly in small localised regions on the separator away from its ends. Each event is very weak and short-lived. These events do not start until $t=0.17t_f$, some $0.08t_f$ after the end of the first phase. During this time we suspect that the system is again trying to regain equilibrium and so builds up the current along the separator. The instant the current anywhere along the separator rises above $j_{crit}$ a little burst of reconnection dissipates it. 

We have found that regions along the separator where the reconnection is strongest are associated with O-type perpendicular magnetic field lines local to the separator and that regions where the reconnection is weaker, in phase I, and throughout all of phase II, are associated with perpendicular magnetic field that is locally X-type. Note that the signature of O-type field actually comes from field lines that twist through just $180^\circ$ as they run up the separator. On the other hand the X-type arises because the newly reconnected field lines slightly twist back on themselves as they run up the separator. Therefore, identifying a separator or separator reconnection from knowledge of the local magnetic field behaviour is not possible, as mentioned in the introduction: separators are global features and therefore knowledge of the global field is required to locate them. However, since strong 3D reconnection is associated with a parallel electric field and, hence, (under the conditions of a classical Ohm's law) is associated with a parallel electric current, magnetic fields that locally have an O-type nature in planes perpendicular to them are potential sites for reconnection.

A counter-rotating flow also exists at a point along the separator close to, but not coincident with, the peak reconnection. This counter-rotation of the flow exists simultaneously with both O-type and X-type flow local to the separator. Indeed, the strongest reconnection appears to be associated with a counter-rotating stagnation flow about the separator with the reconnected field moving outwards towards the nulls and away from the separator. The system starts with very weak flow which appears initially to be elliptic perpendicular to the separator, but then turns to a stagnation-type flow, suggesting that the stagnation flow is generated as a result of the reconnection.


In a companion paper \citep{Stevenson15_jgrb}, we discuss the properties of the waves which are launched from the diffusion site as a consequence of the loss of force balance due to the reconnection. The flows which these waves set up are also discussed in detail.

\begin{acknowledgments}
JEHS undertook much of this work during her PhD during which she was supported financially by STFC. JEHS is now a postdoc in the SMTG in St Andrews funded on their STFC consolidated grant. The computations for this paper were carried out on the St Andrews's UKMHD consortium cluster which is funded by STFC and SRIF. Data from simulation results are available on request from J. E. H. Stevenson (email: jm686@st-andrews.ac.uk).
\end{acknowledgments}







\providecommand{\noopsort}[1]{}\providecommand{\singleletter}[1]{#1}%

\end{article}

\end{document}